# An Empirical Exploration on the Supervision of PhD Students Closely Collaborating with Industry

*Eduard Paul Enoiu*
*eduard.paul.enoiu@mdh.se*
*Software Testing Laboratory, IDT, Västerås*

**Abstract:** With an increase of PhD students working in industry, there is a need to understand what factors are influencing supervision for industrial students. This paper aims at exploring the challenges and good approaches to supervision of industrial PhD students. Data was collected through semi-structured interviews of six PhD students and supervisors with experience in PhD studies at several organizations in the embedded software industry in Sweden. The data was anonymized and it was analyzed by means of thematic analysis. The results indicate that there are many challenges and opportunities to improve the supervision of industrial PhD students.

1. Introduction

In this paper we discuss how the supervision of PhD students is influenced when performing research with industry in software verification and validation. Software testing and verification methods [10] are two of the biggest research directions in software engineering.

The current trend towards empirical research, requires certain collaboration patterns to be established which are guiding the PhD supervision efforts. This paper examines the challenges and perceived good approaches experienced by several PhD students and supervisors. A finding of the research was that while Phd supervision is enhanced by the close collaboration with industry, challenges are posing threats to an efficient and effective supervision.

The discussion which follows is based on a thematic analysis of seven interviews with supervisors and doctoral students from several research groups at Mälardalen University. The next sections will discuss the research questions and methodological approach used in this paper. This will be followed by a section which presents the analysis and findings. Following this, will be a short discussion and then a conclusion.

## 2. Research Questions

There is a lack of evidence on how supervision of PhD students in computer science research can be affected when performing close collaboration with industry. In this paper we explore the challenges and good approaches to supervision of PhD students working in industry. We use several factors to perform thematic analysis. We interviewed seven PhD students and supervisors working with Mälardalen University. We analyzed the interview transcripts with thematic analysis. The following research questions (RQs) were studied:

- RQ1: What challenges supervisors and PhD students perceive when doing research in close collaboration with industry?
- RQ2: What advantages are encountered by supervisors and PhD students when doing research in close collaboration with industry?

## 3. Method

Based on the ethical guidelines proposed by the Swedish Research Council and the Centre of Research Ethics and Bioethics at Uppsala University [1] we took several steps to ensure we fulfill ethical responsibilities. In particular, we anonymized the data and kept strict restrictions in space and time on the raw transcription files, and anonymized the transcripts. Further, we ensured that informed consent was obtained for all participants.

We recruited a convenience sample of individuals affiliated with research groups at Mälardalen University performing industrial research. Three supervisors and four students have participated in this study. Interviews were conducted face-to-face and notes were taken during these interviews. The interviewees were given a lot of freedom to express their thoughts and explain topics not covered by the questions posed.

There are many approaches to qualitative data analysis; we decided to follow thematic analysis as described by Braun and Clarke [2]. This method was suitable for the type of data we had, and it allowed for one sentence of the transcript to be coded as belonging to several themes.

Themes were used as overall topics affecting PhD supervision in general and collaboration with industry specifically. Therefore, we used the topics covered by the Supervisors – Third Cycle Programmes course topics and content stated in the study plan. Completed interviews were added into an online spreadsheet. A list of themes, challenges and good approaches found were used to get consistency in the merged interviews.

## 4. Results

We interviewed seven participants, three supervisors and four PhD students. N this section we describe the PhD supervision process from the perspective of challenges and perceived good approaches during industrial PhD studies. The results are shown in Table 1 and 2. This is based on a synthesis of the information provided by the interviewees and represents an abstract and inclusive view of the PhD supervision factors affecting an industrial oriented PhD project.

All identified challenges and approaches were categorized in the following themes:
1. *Knowledge and Skills in Supervising.* Many factors are affecting the overall process of supervising PhD students. During this process, there is a need to have formal instruction and monitoring processes for developing PhD students skills. Whitelock et al. [3] found that informal collaboration, reflection and creativity are also important during the supervision of PhD students.
2. *Individual Study Plan (ISP).* Another important formal aspect at Mälardalen University is related to the ISP [4]. This is a live document updated regularly that is used for systematically plan and assess the posed quantitative and qualitative in the form of a time plan. In an ISP, a PhD student and its supervisor(s) are supposed to add relevant activities during the progress of the PhD project.
3. *Juridical Regulation and Research Ethics.* Phd students are learning ethical regulations and guidelines as part of their development [5] [6]. This is a factor that is influencing the quality of the supervisory practices.
4. *Equality and Equity in Supervision.* Morley et al. [7] and Lee [8] have explored the area of equality in PhD student supervision in terms of the professional expertise needed and how to improve it in time.
5. *Cooperation, Co-production and Supervision.* Recently, in computer science and other areas of researcher, PhD studentship has evolved to strengthen the collaboration between universities and industry [9] through industrial PhD student projects targeting collaborative researcher. Different funding agencies provide funding for such schemes. Nevertheless, many challenges have been identified [8] with respect to the interaction experience and results of the student, supervisor and industry interaction.

We present the findings related to ten challenges strongly related with the identified themes in PhD supervision (also shown in Table 1). Most of the challenges mentioned by our participants related to the knowledge and skills in supervising and cooperation, Co-production and supervision. For example, PhD students had problems with the willingness of supervisors to listen to the students opinions and with a lack of an effective channel for communication with their supervisors. Supervisors stated also the difficulty of their PhD students to identify research problems and formulating research questions. Related to industrial co-production and collaboration, participants mentioned that balancing industrial expectations and research expectations is difficult. In addition, the lack of feedback from industrial partners and the companies' false expectations towards PhD students and their supervision seem to be a challenge to PhD supervision. Another participant, identified a challenge related to the lack of support from the company in following the ISP during the student's project.

Table 1. Challenges faced during PhD supervision.

| Themes | Challenges |
| --- | --- |
| Knowledge and Skills in Supervising | C1. identifying if the problems are in fact research problems or engineering tasks.<br>C2. difficulty in formulating research questions or challenges during supervision meetings.<br>C3. lack of the willingness of supervisors to listen to their PhD students.<br>C4. lack of an effective and efficient channel for communication between supervisor and PhD student. |
| Local Routines and the Individual Study Plan | C5. lack of support from the company as a PhD student in following the ISP. |
| Juridical Regulation and Research Ethics | C6. supervisors have a lack of clarity on ethical expectations when performing research with human software engineers. |
| Equality and Equity in Supervision | C7. Poor gender equality in supervision and mentoring teams at university and companies. |
| Cooperation, Co-production and Supervision | C8. balancing industrial expectations vs. research expectations<br>C9. lack of feedback from industrial partners.<br>C10. the companies have false expectations towards PhD students and their supervision. |

In addition, we present the findings related to seven perceived good approaches related with the identified themes in PhD supervision (also shown in Table 2). Most of the positive approaches mentioned by our participants related to cooperation, co-production and supervision. It seems that having multiple supervisors is useful during early supervision of PhD students. Several participants have mentioned that the advice from supervisors on working with realistic problems on real data from industry is very helpful. Another participant has mentioned that the ISP can be used for driving the planning meetings with their industrial partners.

Table 2. Good approaches faced during PhD supervision.

| Themes | Good Approaches |
|---|---|
| Knowledge and Skills in Supervising | A1. Having multiple supervisors is useful during early supervision.<br>A2. Advice from supervisors on working with realistic problems on real data from industry is helpful. |
| Local Routines and the Individual Study Plan | A3. ISP can be used for driving the meetings with managers at companies working with PhD students. |
| Cooperation, Co-production and Supervision | A4. Natural bidirectional knowledge transfer between supervisor and PhD student.<br>A5. Getting to know the engineers and how they work in industry and not only working with PhD supervisors.<br>A6. Access to data for evaluation for PhD students during supervision is important and helps.<br>A7. Good access to resources (or authorization to company premises) for both supervisor and supervisee is useful. |

Overall, the results of the thematic analysis show that supervision of PhD students in close collaboration with industry is difficult and poses several challenges related to the (i) knowledge and skills in supervising, (ii) local routines followed and the individual study plan, (iii) the juridical regulation and research ethics, (iv) equality and equity in supervision and also the (v) cooperation and co-production. For example, balancing industrial with research expectations combined with the lack of feedback from industrial partners can negatively influence the quality of the supervision. In addition, the companies have false expectations towards PhD students and their supervision. On the other hand, several positive approaches for good supervision in cooperation with industry have been identified: a natural bidirectional knowledge transfer between supervisor and PhD student, getting to know the engineers in companies and how they work in industry and not only working with PhD supervisors, access to data for evaluation for PhD students during supervision is important and helps and good access to resources (or authorization to company premises) for both supervisor and supervisee is useful.

**5. Conclusions**
Based on the findings of this exploratory interview study, we identified several negative and positive aspects influencing the supervision of PhD students in close collaboration with industry. These aspects can influence future research on this subject, but also can be used by supervisors and PhD students for their reflection. It seems that collaborating with industry during the PhD studies influences the supervision patterns since the students have a different and highly skewed view of how supervision is performed. It seems that supervisors can have a lack of clarity on ethical expectations when performing research with human software engineers. In addition, participants mentioned that there is a poor gender equality in supervision and mentoring teams at university and companies.